\documentclass[letterpaper,aps,prl,superscriptaddress,twocolumn,showpacs]{revtex4}
\usepackage{amsmath}
\usepackage{amssymb}
\usepackage{amsfonts}
\usepackage{txfonts}
\usepackage{courier}
\usepackage{relsize}

\usepackage[T1]{fontenc}
\usepackage{bm}
\usepackage{color}
\usepackage{graphicx}

\newcommand{\g}[1]{\mbox{\boldmath $#1$}}
\newcommand{\Eq}[1]{Eq.~\eqref{#1}}
\newcommand{\Fig}[1]{Fig.~\ref{#1}}
\newcommand{\Ref}[1]{Ref.~\cite{#1}}

\newcommand{\blue}[1]{#1}%{\textbf{\textcolor{blue}{#1}}}%{#1}%

\newcommand{\thistext}{Letter} %letter or paper, depending on journal

\begin{document}

\title{Effective Critical Electric Field for Runaway-Electron Generation}

\author{A. Stahl}
\email[]{stahla@chalmers.se}
\affiliation{Department of Applied Physics, Chalmers University of Technology, SE-412 96 Gothenburg, Sweden}
\author{E. Hirvijoki}
\affiliation{Department of Applied Physics, Chalmers University of Technology, SE-412 96 Gothenburg, Sweden}
\author{J. Decker}
\affiliation{\'Ecole Polytechnique F\'ed\'erale de Lausanne (EPFL), Centre de Recherches en Physique des Plasmas (CRPP),\\ CH-1015 Lausanne, Switzerland}
\affiliation{Department of Applied Physics, Chalmers University of Technology, SE-412 96 Gothenburg, Sweden}
\author{O. Embr\'eus}
\affiliation{Department of Applied Physics, Chalmers University of Technology, SE-412 96 Gothenburg, Sweden}
\author{T. F\"{u}l\"{o}p}
\affiliation{Department of Applied Physics, Chalmers University of Technology, SE-412 96 Gothenburg, Sweden}

\date{\today}

\begin{abstract}  
In this \thistext\ we investigate factors that influence the effective critical electric field for runaway-electron generation in plasmas. We present numerical solutions of the kinetic equation and discuss the implications for the threshold electric field. We show that the effective electric field necessary for significant runaway-electron formation often is higher than previously calculated due to both (1) extremely strong dependence of primary generation on temperature, and (2) synchrotron radiation losses. We also address the effective critical field in the context of a transition from runaway growth to decay. We find agreement with recent experiments, but show that the observation of an elevated effective critical field can mainly be attributed to changes in the momentum-space distribution of runaways, and only to a lesser extent to a de facto change in the critical field.
\end{abstract} 


\maketitle %\maketitle must follow title, authors, abstract and \pacs

%%%%%%%%%%%%%%%%%%%%%%%%%%%%%%%%%%%%%%%%%%%%%%%%%%%%%%%%%%%
{\em Introduction.}---In a plasma, an electron beam accelerated by an electric field
is damped by collisional friction against the bulk plasma and by
emission of electromagnetic radiation. 
Since the collisional friction decreases with increasing
velocity of the electrons, a large enough electric field may overcome the collisional damping and accelerate 
electrons to relativistic speeds, leading to the formation of a runaway-electron (RE) beam. 
In laboratory plasmas, much attention has been given to the potentially dangerous, 
highly relativistic RE beams that can be generated in tokamak disruptions \cite{helander}. 
Runaway acceleration can also occur in
nondisruptive plasmas due to the Ohmic electric field, if the plasma
density is low.  In addition, runaway electrons are
ubiquitous in atmospheric and space plasmas, e.g., as a source of red sprites in the mesosphere \cite{bell} and in lightning discharges in thunderstorms \cite{gurevich}, and their occurrence in solar flares has been suggested \cite{holman}. 

The critical (threshold) electric field for runaway electron
generation, $E_c=n_ee^3\ln{\Lambda} /4\pi\epsilon_0^2m_ec^2$, is a
classic result in plasma physics~\cite{connor-hastie}. Because of
relativistic effects, it is the weakest field at which electron
runaway is possible. Here, $n_e$ and $m_e$ are the number density and rest
mass of the electrons, respectively, $\ln{\Lambda}$ is the Coulomb
logarithm, $c$ is the speed of light, and $e$ is the magnitude of the
elementary charge. Recent experimental evidence from several tokamaks
indicates that the electric field strength necessary for RE generation
could in fact be several times larger than the critical electric
field~\cite{granetz,Paz-Soldan,ms}. As the classic expression for the
critical field only considers the balance between electric field and
Coulomb collisions, many potential mechanisms affecting the RE
generation are left out. In this \thistext\ we investigate the role of
the background plasma temperature and synchrotron radiation reaction
as possible explanations for these observations.

%%%%%%%%%%%%%%%%%%%%%%%
%This figure was obtained using plot_E_over_Ec.m 
\begin{figure}
\includegraphics[width=0.45\textwidth]{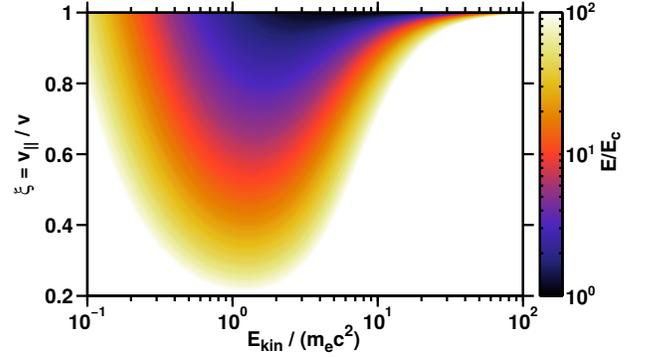}
\caption{Magnitude of the normalized electric field ($E/E_c$) necessary to compensate for collisional friction and synchrotron damping for a single electron in an ITER-like plasma with $n_e=10^{20}$~m$^{-3}$, $T_e=10$~keV, and $B=5$~T.}
\label{fig:single_particle_energy_balance}
\end{figure}
%%%%%%%%%%%%%%%%%%%%%%%
 
An appreciation for the importance of temperature and synchrotron effects can be gained by considering the energy balance for an electron experiencing electric field acceleration, collisional damping, and the Abraham-Lorentz radiation reaction force \cite{al}
\begin{equation}
\mathbf{F}_{\textrm{rad}}=k\left[\ddot{\mathbf{v}}+\frac{3\gamma^{2}}{c^{2}}\left(\mathbf{v}\cdot\dot{\mathbf{v}}\right)\dot{\mathbf{v}}+\frac{\gamma^{2}}{c^{2}}\left(\mathbf{v}\cdot\ddot{\mathbf{v}}+\frac{3\gamma^{2}}{c^{2}}\left(\mathbf{v}\cdot\dot{\mathbf{v}}\right)^{2}\right)\mathbf{v}\right], 
\label{eq:F_rad}
\end{equation}
where $k=e^{2}\gamma^{2}/(6\pi\varepsilon_{0}c^{3})$, $\mathbf{v}$ is the electron velocity, and \mbox{$\gamma=[1-(v/c)^2]^{-1/2}$} is the relativistic mass factor. At the critical electric field, acceleration
due to the electric field balances the friction due to collisions and
radiation, so the particle energy is constant. This means that $\dot{\gamma} = 0$ and $\g v \cdot \dot{\g v}=0$, implying that $q \g
E\cdot \g v +\g F_c\cdot \g v -(e^2 \gamma^4/6\pi\epsilon_0
c^2)\dot{\g v}\cdot\dot{\g v}=0$. At constant energy, $\dot{\g
  v}=(e/\gamma m_e) \g v \times \g B$,  so that $\dot{\g v}\cdot\dot{\g
  v}=(1-\xi^2)v^2\omega_c^2$. Here $\g B$ is the magnetic field, $\xi=p_\parallel/p$ is the cosine of the particle pitch angle, $p\!=\!|\mathbf{p}|\!=\!\gamma v/c$ is the magnitude of the normalized momentum, and $\omega_c=e
B/\gamma m_e$. This means that for the electric field to accelerate
the electron, it has to be larger than
\begin{equation}
\frac{E}{E_c}>\frac{1}{\xi}\frac{c^2}{v^2}+\frac{2\varepsilon_0B^2}{3n_em_e\ln\Lambda}\frac{1-\xi^2}{\xi}\frac{v}{c}\gamma^2 \equiv h(v,\xi).
\label{eq:E/E_c}
\end{equation}
Note that $h(v,\xi)> 1$, always. In \Fig{fig:single_particle_energy_balance}, $h$ is illustrated for typical ITER \cite{iter} plasma parameters as a function of $\xi$ and the electron kinetic energy~$E_{\rm{kin}}$. From this simple estimate of the single electron energy balance, we see that the value of $E/E_c$ needed to accelerate electrons 
can be significantly larger than unity and increases with $E_{\rm{kin}}$ in the MeV range. 
This warrants a more thorough investigation of the RE dynamics close to the critical electric field. The first term in \Eq{eq:E/E_c} is related to the collisional damping, which is temperature dependent, while the second term--which vanishes for purely parallel motion ($\xi=1$)--is due to the radiation reaction force. We will consider both these effects in more detail. 

The single electron estimate depends strongly on $\xi$, $v$, and various plasma parameters, but neglects the collisional Coulomb diffusion that spreads the electrons in velocity space. An accurate estimate for the
threshold electric field can thus only be obtained using kinetic
calculations that take into account the details of the electron
distribution function. Here we make use of COllisional Distribution of Electrons (CODE)~\cite{CODE_paper}, an efficient
finite-difference--spectral-method tool that solves the
two-dimensional momentum space kinetic equation in a homogeneous
plasma. The Coulomb collision operator in CODE is valid for arbitrary
electron energies~\cite{papp2011}. Often secondary (or {\em avalanche}) generation of REs, resulting from knockon collisions between REs and thermal electrons, is the dominant RE generation mechanism, and CODE has also been
equipped with an operator describing this process~\cite{rosput}. For the present work, an operator for synchrotron emission backreaction based on the analysis in Refs.~\cite{andersson01,hazeltine} was implemented. With
the numerical electron distribution function from CODE, we may
investigate the runaway generation dynamics for a wide range of plasma
parameters, to calculate, e.g., the synchrotron radiation spectra of the
REs~\cite{syrup_paper}, or to study wave-particle
interactions~\cite{whistler,nearcrit}. The parameters used in this
\thistext\ reflect those common to magnetic fusion experiments, but
the arguments are generally applicable. In particular, no effects specific to fusion plasmas (such as a toroidal field configuration) have been assumed.

%%%%%%%%%%%%%%%%%%%%%%%%%%%%%%%%%%%%%%%%%%%%%%%%%%%%%%%%%%%%%%%%%%%%%%
{\em Temperature dependence of the critical electric field.}---At $E\!\gtrsim\! E_c$, only electrons already moving with approximately
the speed of light may run away.  Since the number of plasma particles is finite (especially in a laboratory context), the actual highest speed
achieved by the background electrons may be significantly less than
$c$. Thus, if the critical speed for RE generation at a given
$E$ field is larger than this maximum speed, no electrons will be able
to run away. The width (in velocity space) of the 
distribution function describing the particle speeds is determined by the
temperature. This introduces a temperature dependence to the
\emph{effective} critical field, since the number of particles with
speed above any threshold speed is temperature dependent.
Mathematically, this can be understood from the primary runaway growth
rate \cite{connor-hastie}:
\begin{align*}
\frac{d n_r}{dt}\sim  n_e \nu_{ee}\, {\mathcal E}^{-3 (1+Z_{\rm eff})/16}\exp{\left[-1/(4\mathcal E)-\sqrt{(1+Z_{\rm eff})/\mathcal E}\right]},
\end{align*}
(where $\nu_{ee} = n_e e^4
\ln\Lambda/4\pi\varepsilon_0^2m_e^2v_{\mbox{\scriptsize{th}}}^3$ is the collision
frequency, $v_{\mbox{\scriptsize{th}}}\!=\!\sqrt{2T_e/m_e}$ is the electron thermal velocity, and $Z_{\rm{eff}}$ is the effective charge number of the plasma), which
is exponentially small in $\mathcal E= E/E_D = (T_e/m_e c^2)
(E/E_c)$, where $E_D$ is the Dreicer field \cite{dreicer}. There is thus an inherent temperature dependence in the primary runaway growth rate at a given value of $E/E_c$, and for significant RE production on a short time scale it is not enough to only require~$E>E_c$ \blue{\cite{jayakumar,helander}}.

\begin{figure}
\includegraphics[width=0.47\textwidth]{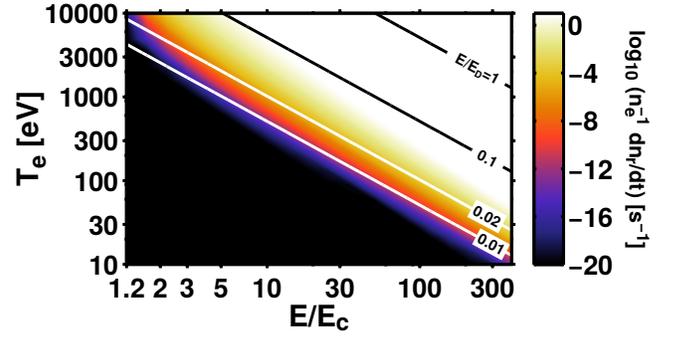}
\caption{Primary runaway growth rate (particle fraction per second) as a function of temperature and electric field, in the absence of synchrotron effects. White and black contours refer to
  $E/E_D$. The plasma parameters $n_e=5\times 10^{19}$~m$^{-3}$ and
  $Z_{\rm{eff}}=1.5$ were used.}
\label{fig:temp_dist}
\end{figure}

We define a runaway electron as any electron with $p\!>\!p_c\!=\!(E/E_c \!-\!1)^{-1/2}$, where $p_c$ is the critical momentum for electron runaway. In the absence of avalanche generation, a quasisteady state for the RE distribution can be calculated using CODE. In~\Fig{fig:temp_dist}, the RE growth rate for this primary distribution is displayed as a function of the electron temperature and
$E/E_c$. The figure indicates that, for all temperatures $T_e\!\lesssim\! 5$~keV, the fraction of the electron population that runs away in 1~s is less than $10^{-20}$ for all $E/E_c < 1.5$. In a plasma with $n_e\lesssim
10^{20}\,\textrm{m}^{-3}$ and a volume of a few tens of m$^3$ (typical
of fusion experiments), essentially no runaway production (let alone
detection) is thus to be expected. It is also clear from the figure that, for lower
temperatures, a stronger normalized electric field would be required for
significant RE production (note that \Fig{fig:temp_dist} essentially
covers the whole temperature range of magnetic fusion plasma
operation). We note that this temperature dependence may increase the sensitivity of future hotter tokamaks (like ITER) to the problem of deleterious RE formation. The white and black contours in \Fig{fig:temp_dist} show
the corresponding values of $E/E_D$, and we may conclude that $E/E_D$
must be at least larger than $1\%-2\%$ for significant runaway formation
to occur, in agreement with previous analytical findings \cite{jayakumar,helander}. In practice, there are thus two conditions that must be
fulfilled: $E/E_c>1$ and $E/E_D> k$, for some $k\!\sim\! 1\%\!-\!2\%$. The second criterion is more restrictive for temperatures below 5.1 and 10.2 keV (for $k=1\%$ and $2\%$), respectively. 

%%%%%%%%%%%%%%%%%%%%%%%%%%%%%%%%%%%%%%%%%%%%%%%%%%%%%%%%%%%%%%%%%%%
%%% This figure was obtained using SynchrotronLossFigures.m 
%%%%%%%%%%%%%%%%%%%%%%%%%%%%%%%%%%%%%%%%%%%%%%%%%%%%%%%%%%%%%%%%%%%

%%%%%%%%%%%%%%%%%%%%%%%%%%%%%%%%%%%%%%%%%%%%%%%%%%%%%%%%%%%%%%%%%%%
%%% This figure was obtained using SynchrotronLossFigures.m 
\begin{figure}
\begin{center}
{\includegraphics[trim=0.0cm 0cm 0cm 0cm, clip=true, width=0.41\textwidth]{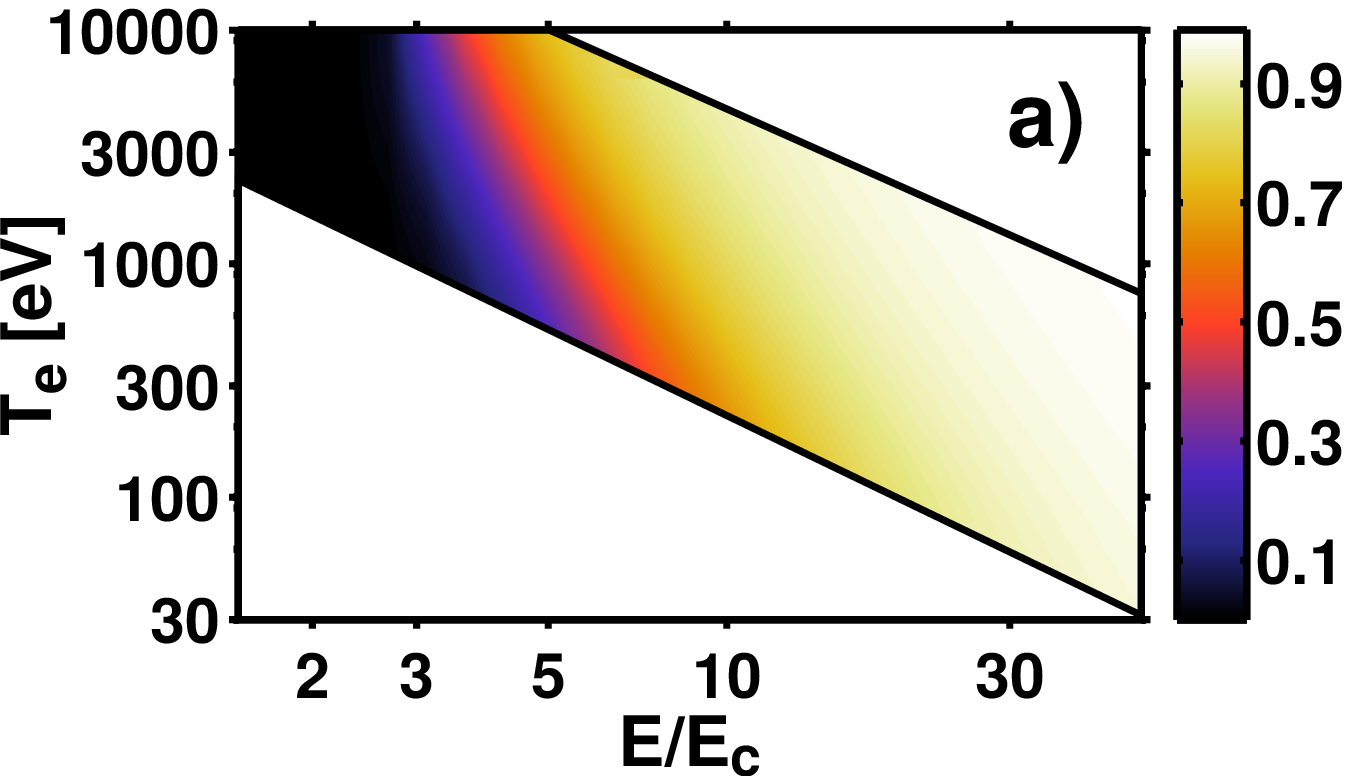}}\hfill
{\includegraphics[trim=-0.0cm 0cm 0cm 0cm, clip=true, width=0.4\textwidth]{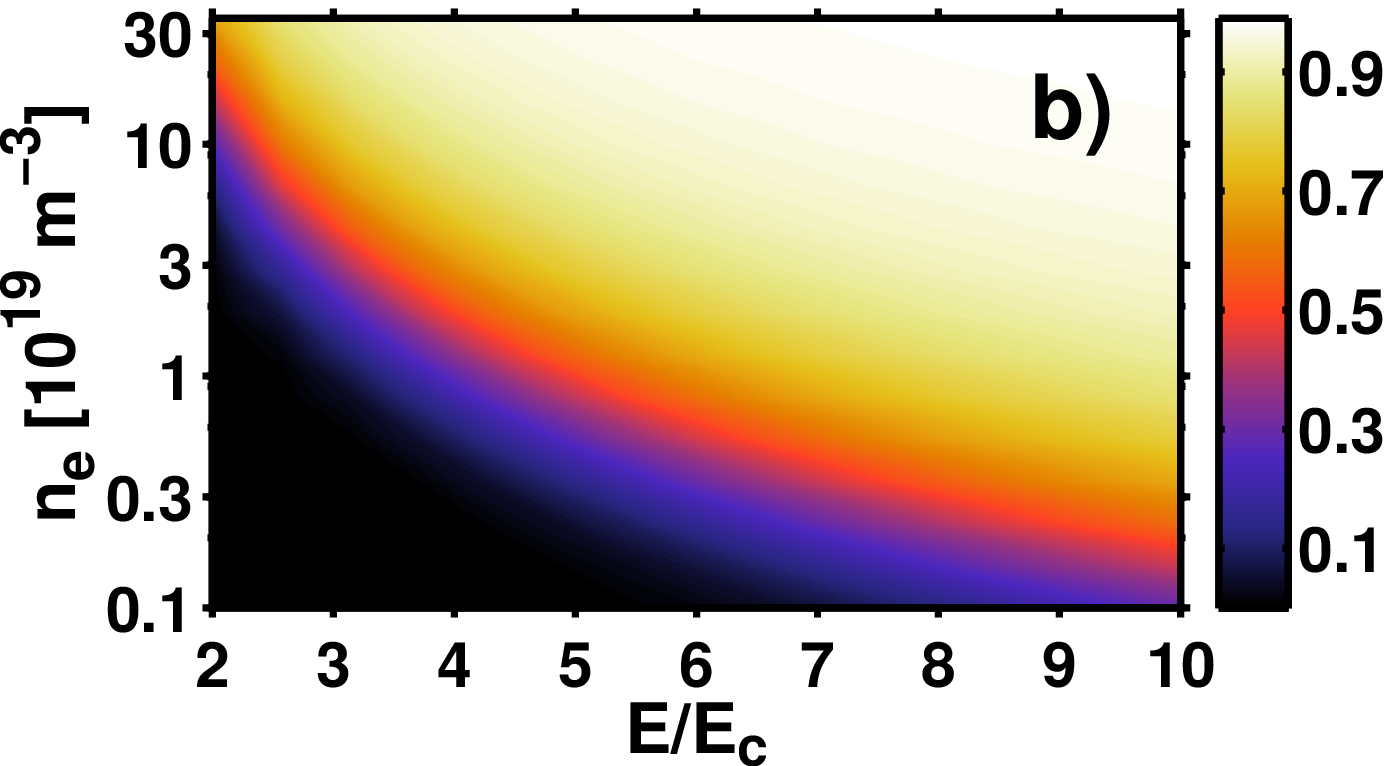}}
\caption{Contour plots of the (a) temperature and (b) density dependence of the ratio between the primary RE growth rate in CODE with and without synchrotron effects included. The parameters \mbox{$B=4$~T}, $Z_{\rm{eff}}=1.5$, and (a) $n_e=1\times 10^{19}$~m$^{-3}$, (b) $T_e=2$~keV were used. To ensure reliable results in (a), the parameter region has been restricted as the growth rates are negligible for low $T_e$ and $E$ fields, and $E/E_D$ approaches unity for high $T_e$ and $E$ fields (cf. \Fig{fig:temp_dist}).}
\label{fig:synch}
\end{center}
\end{figure}
%%%%%%%%%%%%%%%%%%%%%%%%%%%%%%%%%%%%%%%%%%%%%%%%%%%%%%%%%%%%%%%%%%%

%%%%%%%%%%%%%%%%%%%%%%%%%%%%%%%%%%%%%%%%%%%%%%%%%%%%%%%%%%%
{\em Momentum loss due to synchrotron emission.}--- 
\blue{The importance of synchrotron backreaction as a limiting factor for the maximum energy achieved by REs has been discussed before \cite{ms98}, and its importance in RE dynamics has been investigated \cite{andersson01}, also in the context of the critical field for RE generation \cite{ms}. An accurate description of the RE dynamics close to the critical field based on first principles does, however, require kinetic modeling, and we will investigate the effect of the synchrotron emission on the effective critical field using CODE.}  

In a homogeneous plasma, the distribution function $f$ for electrons experiencing an electric field, Coulomb collisions, and synchrotron
radiation is determined by the gyro-averaged Fokker-Planck equation,
\begin{equation*}
\frac{\partial f }{\partial t}+\underbrace{\frac{e E_{\parallel}}{m_e c}\left(\xi
\frac{\partial f}{\partial p}+\frac{1-\xi^2}{p}\frac{\partial f}{\partial \xi}\right)}_{\mbox{electric field}} +\underbrace{\frac{\partial}{\partial\mathbf{p}}\cdot\left(\mathbf{F}_{\textrm{rad}}f\right)}_{\mbox{radiation}} =\, C\{f\} + S_{\!\mbox{ava}},
\label{fokkerplanck}
\end{equation*}
where $E_{\parallel}$ is the parallel (to $-\mathbf{B}$) electric field,
$C\{\cdot\}$ is the collision operator, $S_{\!\mbox{ava}}$ is the
source of secondary (avalanche) runaways, and
$\mathbf{F}_{\textrm{rad}}$ is given by \Eq{eq:F_rad}. Note that the operator $\displaystyle (\partial/\partial\mathbf{p})\cdot\left(\mathbf{F}_{\textrm{rad}}f\right)$ conserves the number of particles, unlike the corresponding operator used in \Ref{andersson01}. (The simplified operator used in \Ref{andersson01} was justified due to the focus on the high-energy tail of the distribution function.) The magnetic
force $|\mathbf{F}_m|\simeq \omega_c mv$, characterized by the Larmor
frequency $\omega_c=qB/\gamma m$, typically dominates both the
electric force and the radiation reaction force. This implies that
$\mathbf{v}\cdot\dot{\mathbf{v}}\simeq 0$, and that in the coordinates
$p$ and $\xi$, the term accounting for the effects of synchrotron radiation 
can be written as
\begin{equation*}
\frac{\partial}{\partial\mathbf{p}}\cdot\left(\mathbf{F}_{\textrm{rad}}f\right)=-\frac{1}{p^2}\frac{\partial}{\partial p}\left(\frac{\gamma p^3(1-\xi^2)}{\tau_r}f\right)+\frac{\partial}{\partial\xi}\left(\frac{\xi (1-\xi^2)}{\gamma\tau_r}f\right),
 \label{eq:lossterm}
 \end{equation*}
where $\tau_r=6\pi\varepsilon_0(m_{e}c)^{3}/(e^{4}B^{2})$ \cite{andersson01} is the radiation damping time scale. 
What ultimately determines the relative importance of the synchrotron effects is the ratio of the collision time $1/\nu_{ee}$ to the radiation time scale~$\tau_r$. For a given magnetic field, we therefore expect the largest effect on the distribution for high temperatures and low densities, since $1/(\nu_{ee} \tau_r)\sim T_e^{3/2}B^2/n_e$.

The radiation reaction force acts as an additional drag, which increases with particle momentum. Therefore, it ultimately prevents the REs from reaching arbitrary energies \blue{\cite{ms98}}, and given enough time, the system will reach a steady state where the RE growth rate vanishes. This occurs only once the REs have reached very high energies; in the initial phase (which is our interest in this section), the RE growth rate is well defined. The rate is calculated as the flux through a sphere of constant $p$, located well inside the RE region. The change in this RE growth rate in CODE as a result of the synchrotron radiation reaction is presented in \Fig{fig:synch}.
From the figure, we conclude that the synchrotron losses can reduce the RE rate substantially for weak $E$ fields--by several orders of magnitude at high temperatures and low densities--and it is therefore essential to include these effects when considering near-critical RE dynamics. The sharp cutoff for weak fields is in line with the change in effective critical field associated with the inclusion of the synchrotron drag [see Eq.~\eqref{eq:E/E_c}]. The full kinetic simulation thus agrees qualitatively with the single-particle estimate in the Introduction. For stronger electric fields, the effects are less pronounced.  

We note that in the post-thermal-quench conditions associated with disruptions in tokamaks (low $T_e$, high $n_e$), the effects of synchrotron radiation reaction on the RE growth rate are likely to be negligible, whereas in the case of RE generation during the plasma current ramp-up phase (high $T_e$, low $n_e$) they can be substantial. There is thus a
qualitative difference in the momentum space dynamics in these two
cases--at least for near-critical electric fields--and conclusions from ramp-up (or flattop) scenarios do not necessarily apply under post\-disruption conditions.

%%%%%%%%%%%%%%%%%%%%%%%%%%%%%%%%%%%%%%%%%%%%%%%%%%%%%
{\em Critical field under experimental conditions.}---Until
now we have only considered primary (Dreicer \cite{dreicer}) generation, i.e., when
the electrons gradually diffuse through velocity space due to
small-angle collisions and run away as they reach the critical
velocity. An electron can enter the runaway region also through a
sudden collision at close range, which throws it above the critical
speed in a single event. 
This leads to avalanche multiplication of the REs, which is the dominant mechanism in many cases. The avalanche growth rate is
$\gamma_{\mbox{ava}} \propto n_{\mbox{\scriptsize{RE}}}(E/E_c-1)$ \cite{rosput} (where $n_{\mbox{\scriptsize{RE}}}$ is the runaway density), and the total runaway
growth rate is then the sum of primary and avalanche processes.

Above the critical field, the RE population will be growing and below it will
be decaying; according to the estimate in Ref.~\cite{rosput} (taking only
losses due to Coulomb collisions into account), the transition should
occur at $E/E_c=1$.
However, in an experiment designed to test this
at the DIII-D tokamak \cite{Paz-Soldan}, the measured transition from RE growth to decay occurred at $E/E_c=3\!-\!5$. In the experiments, after first generating a substantial RE population, the plasma density was rapidly increased so as to raise $E_{c}$. The value of $E/E_{c}$ for the transition was determined
from the change from growth to decay in the hard-x-ray (HXR) signal, as well as from the synchrotron emission (in the visual range). These signals were thus treated as a straightforward representation of the number of REs. This assumption is not necessarily valid in the present context, as it neglects the influence of the distribution of RE energies on the emitted radiation. We illustrate this by calculating the synchrotron emission from RE distributions obtained with CODE.

The synchrotron radiation emitted by an electron is highly dependent on its energy and the curvature of its trajectory, with highly energetic particles with large pitch angles emitting most strongly and at the shortest wavelengths. The total emission is therefore very sensitive to the shape of the electron distribution function \cite{syrup_paper}. A change in the distribution shape may in fact lead to a reduction in the emitted synchrotron power, even though the RE population is constant, or even increasing. Here we use the code SYRUP \cite{syrup_paper} to calculate the synchrotron spectrum emitted by CODE distributions, in order to investigate the role of this effect in the experiment in \Ref{Paz-Soldan}. 

Accessing the physics underlying the evolution of the RE distribution
in the experiment may by done in CODE by ramping
down the electric field strength in the presence of a significant RE
tail, dominated by the avalanche mechanism. The purpose here is to give a qualitative explanation for the observed discrepancy.
The CODE calculation was started with a constant $E/E_c\!=\!12$, to produce a significant RE tail, and after 1.2~s the electric field was gradually ramped down
during 1~s---a time scale consistent with the experiments described in
\Ref{Paz-Soldan}. The synchrotron spectrum at each time step was calculated using SYRUP. In the calculations, the maximum particle
energy was restricted to 22 MeV, in agreement with experimental
observations \cite{hollmann}. (This maximum energy limit cannot be
attributed to the effects of the synchrotron losses; other 
mechanisms, such as radial transport, must be invoked to explain it,
but this is outside the scope of this \thistext.)

\begin{figure}
\begin{center}
{\includegraphics[trim=0cm 0cm 0cm 0cm, clip=true, width=0.48\textwidth]{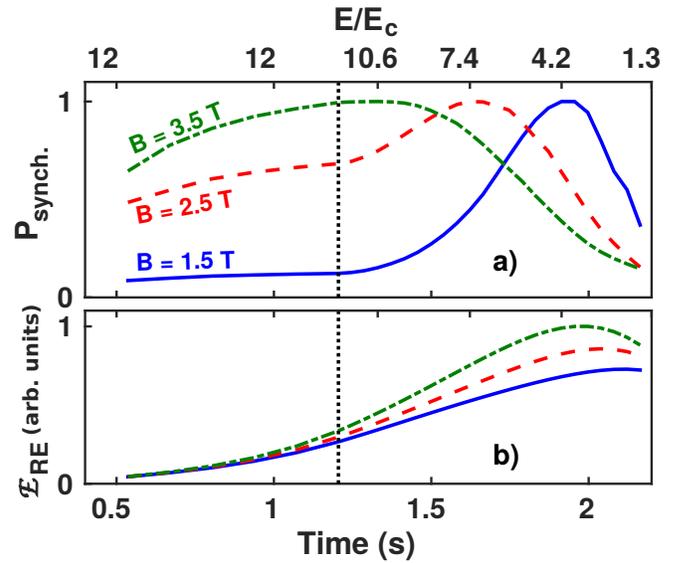}}
\caption{(a) Emitted synchrotron power in the visual range and (b) kinetic energy contained in the RE population during electric field ramp-down. In (a), each curve is normalized to its peak value. The parameters $T_e=1.1$~keV, \mbox{$n_e=2.5\times 10^{19}$~m$^{-3}$} and $Z_{\rm{eff}}=1.2$ were used. The black dotted lines denote the beginning of the $E$-field ramp-down phase.}
\label{fig:growth_to_decay}
\end{center}
\end{figure}

Figure \ref{fig:growth_to_decay}(a) shows the total emitted power in the visual spectral range (400--700 nm) during the simulation, for various $B$-field strengths. The figure shows that as $E/E_c$ decreases, the emitted power transitions from growth to decay in all cases, even though $E/E_c$ is still well above unity. There is a clear dependence on the magnetic field strength, as the transition occurs at $E/E_c\simeq 7.4$ and $10.6$ for $B=2.5$ and $3.5$~T, respectively. This suggests that the origin of the effect is indeed the influence of the synchrotron reaction force on the distribution. In the experiments in \Ref{Paz-Soldan}, the field was 1.5~T; in this case, we observe an apparent effective critical field $E/E_c\simeq 4.2$, in agreement with the experimental value of 3--5.

Although the RE growth rate decreases with $E/E_c$, in the calculations it remains positive for all $E/E_c\!>1.1$, $1.3$, and $1.4$ for $B=1.5$, $2.5$, and $3.5$ T, respectively. The effective critical field is thus close to unity in all cases considered, and is $B$ dependent, as expected from the previous section. The kinetic energy content of the RE population also continues to increase well after the emitted synchrotron power has started to decrease, as shown in \Fig{fig:growth_to_decay}(b). We therefore conclude that what is observed in \Fig{fig:growth_to_decay}(a) is not a fundamental change in the critical electric field but a reduction in synchrotron emission. The changing electric field leads to a reduced accelerating force, which modifies the force balance, causing a redistribution of electrons in velocity space towards lower energies. The density of highly energetic particles with large pitch angles thereby decreases, leading to a substantially reduced synchrotron emission in the visual range.

Trends similar to those shown in \Fig{fig:growth_to_decay}(a) are seen also in the infrared spectral range. Since several tokamaks are equipped with fast visual or IR cameras dedicated to observing the synchrotron emission from REs, experimental study of the effect we describe should be within reach. In particular, confirming the $B$ dependence of the apparent elevated critical field would be of interest.

%%%%%%%%%%%%%%%%%%%%%%%%%%%%%%%%%%%%%%%%%%%%%%%%%%%%%%%%%%%%%%%%%%%
%%% This figure was obtained using SynchrotronFigureForPaper.m 
%%%%%%%%%%%%%%%%%%%%%%%%%%%%%%%%%%%%%%%%%%%%%%%%%%%%%%%%%%%%%%%%%%%

In the experiments, $E/E_c$ was decreased via a density ramp-up. Increasing the density also indirectly modifies $T_e$, $Z_{\rm{eff}}$, and the loop voltage \cite{Paz-Soldan} and reduces the magnitude of the synchrotron effects. In the calculation in \Fig{fig:growth_to_decay} these changes were not taken into account; however, the plasma parameters where chosen to reflect those observed at the time of the transition from growth to decay in DIII-D shot 153545, considered in \Ref{Paz-Soldan}. Although the trend shown in \Fig{fig:growth_to_decay} appears consistently and the results are largely independent of the details of the ramp in electric field strength and the energy cutoff, the specific value of $E/E_c$ for which the transition from growth to decay occurs is sensitive to the plasma parameters. \blue{Similarly, the details of the avalanche source used \cite{rosput} are expected to affect only the specific value of the transition, not the qualitative behavior in \Fig{fig:growth_to_decay} (indeed, the same trend is seen even when the avalanche process is not included at all).}

It was suggested in \Ref{Paz-Soldan} that a significant part of the detected HXR signal was due to RE bremsstrahlung emission. Like the synchrotron emission, bremsstrahlung is sensitive to the RE distribution function---the general argument made above may be invoked to explain the elevated growth-to-decay transition also in the case of the HXR signal. The effects considered in this section thus offer a plausible explanation for the mechanism behind the experimentally detected elevated electric field for transition from RE growth to decay. Our simulations show that the observed increase is mainly an artifact of the methods used to determine it, and only to a lesser extent the result of a fundamental change to the critical field itself. The impact on runaway mitigation schemes in future tokamaks is likely to be negligible, especially considering that in a postdisruption scenario the impact on the distribution function from synchrotron backreaction is small (due to the low $T_e$ and high $n_e$).

%%%%%%%%%%%%%%%%%%%%%%%%%%%%%%%%%%%%%%%%%%%%%%%%%%%%%%%%%%%
{\em Conclusions.}---We have shown that several factors can influence
the effective critical electric field for both generation and decay of runaway
electrons. The temperature dependence of the RE growth rate means that
in practice, $E/E_D>$1\%--2\% is required for substantial RE
generation. In addition, the drag due to synchrotron emission backreaction increases the critical field; for weak $E$ fields, the
runaway growth rate can be reduced by orders of magnitude. The synchrotron effects on the distribution are most prominent at high temperature and low density, however, and their practical impact is likely negligible in postdisruption tokamak plasmas. By the same token, the effects can be substantial during ramp-up and flattop. 

Deducing changes to the size of the runaway population using radiation can be misleading, as the emission is very sensitive to the momentum-space distribution of runaways. This can lead to a perceived elevated critical field in electric field ramp-down scenarios, despite a continuing increase in the energy carried by the runaways and their density. Our results are consistent with recent experimental observations, giving a possible explanation for the observed elevated critical field.

%%% If you have acknowledgments, this puts in the proper section head.
\begin{acknowledgments}
The authors are grateful to M. Landreman, G. Papp, G. Pokol, and I. Pusztai for fruitful
discussions. This work has been carried out within the framework of the EUROfusion
Consortium and has received funding from the Euratom research and training program
2014-2018 under Grant Agreement No. 633053. The views and opinions expressed herein
do not necessarily reflect those of the European Commission.
\end{acknowledgments}

% Bibliography
\bibliographystyle{unsrt}

\end{document}